# Intrinsic Spectrum Analysis of Laser Dynamics Based on Fractional Fourier Transform

Ligang Huang,[1,2] Tianyi Lan,[1,2] Chaoze Zhang,[1,2] Laiyang Dang,[1] Tianyu Guan,[1] Bowen Zheng,[1] Shunli Liu,[1] Lei Gao,[1] Wei Huang,[1] Guolu Yin,[1] and Tao Zhu[1,*]

[1]Key Laboratory of Optoelectronic Technology & Systems (Ministry of Education), Chongqing University, Chongqing 400044, China.
[2]These authors contributed equally.
*E-mail: zhutao@cqu.edu.cn

Intrinsic spectrum that results from the coupling of spontaneous emission in a laser cavity, can determine the energy concentration and coherence of lasers, which is crucial for the optical high-precision measurement. Up to now, it is hard to analyze the intrinsic spectrum in the high-speed laser dynamics process, especially under the condition of fast wavelength sweeping. In this work, a new method to analyze the laser intrinsic spectrum is proposed with the laser energy decomposition to a series of chirp-frequency signals, which is realized by fractional Fourier transform (FRFT) of the coherently reconstructed laser waveform. The new understanding of the energy distribution of lasers contributes to the accurate characterization of laser dynamical parameters in time-frequency domain. In the proof-of-concept experiment, the time-frequency dynamical process of a commercial wavelength swept laser is tested with different wavelength-scanning speeds, and the most suitable measurement time window width required for the FRFT-based narrowest spectrum is also explored. The proposed analysis method of laser dynamical parameters will promote the understanding of laser dynamics, and benefit for the optical precision measurement applications.

## 1. Introduction

High-performance lasers with dynamical time-frequency parameters, such as wavelength-swept lasers and chirped ultrashort pulse lasers, have been widely used in the fields of optical precision measurement, such as frequency-modulated continuous-wave (FMCW) Lidar[1], precision spectrum[2,3], optical frequency domain reflectometer (OFDR) distributed fiber sensing[4,5], swept-source optical coherence tomography (SS-OCT)[6], and so on. The time-frequency analysis of lasers is crucial for the performance characterization, including the energy concentration and sweeping speed. For the laser applications, the energy concentration of lasers in time-frequency domain, commonly described by linewidth, characterizes the laser coherence and spectrum purity, which determines the maximum measurement distance and depth of FMCW Lidar, OFDR fiber sensing and SS-OCT. The frequency sweeping speed determines the response time of the optical precision measurement system, and needs to be accurately measured for the frequency comparison. For the operation of dynamical time-frequency laser, the spontaneous emission couples with the laser longitudinal mode back and forth in the laser cavity, and induces the random amplitude and phase modulation, which forms the intrinsic spectrum broadening of the laser[7-9]. Therefore, the dynamical measurement of the intrinsic spectrum induced by spontaneous emission is of great importance to optimize the laser structure for the improvement of the energy concentration and coherence of dynamical time-frequency laser.

Up to now, the concept of energy distribution refers to the frequency-domain linewidth when decomposing the laser waveform into a series of fixed-frequency signals. On this basis, the current measurement methods of dynamical time-frequency laser linewidth mainly include direct spectrum scanning, roll-off measurement and coherent detection. The direct spectrum scanning method obtains the transient spectrum of lasers by scanning a narrow-band filter such as a Fabry-Perot (FP) cavity[10], so as to measure the laser linewidth. The roll-off method[11,12] measures the attenuation relationship between the interference spectrum intensity and the arm length difference of the interferometer, and obtains the coherence length and coherence time of the dynamical time-frequency laser, which can be used to deduce the average linewidth. The coherent detection method mainly includes three techniques of coherent receiver detection[13], delayed self-heterodyne/homodyne detection[14-18] and 3×3 coupler interferometry[19,20], which can reconstruct the laser amplitude and phase. The transient laser linewidth and noise can be further obtained by Fourier transform of the reconstructed optical waveform. However, in the case of high-speed laser frequency scanning, the concept of fixed-frequency decomposing of the laser energy will lead to the broadening of the laser linewidth. For example, when the laser frequency scanning speed reaches 1 MHz/μs, the linewidth broadening induced by the laser scanning will reach the order of MHz in 1 μs time window. Thus, the conventional fixed-frequency energy decomposing cannot reflect the signal purity and noise intensity of the fast swept laser with low noise. It is necessary to redefine the energy decomposition waveform according to the intrinsic feature of frequency sweeping. It is worth noting that the fractional Fourier transform (FRFT) can decompose the wave energy with the frequency-swept waveform as the basic component signal, which has been widely used in optics[21], signal processing[22,23], radar[24], communication[25], and so on.

In this work, we propose to analyze the intrinsic energy distribution of dynamical time-frequency laser, by decomposing the laser energy with chirp-frequency waveform signal as the basic component, which is realized by fractional Fourier transform (FRFT) of the coherently reconstructed laser waveform. In the proof-of-concept experiment, the intrinsic energy distribution and linewidth of a commercial wavelength swept laser as the typical dynamical time-frequency laser are tested with different scanning speeds, and the appropriate measurement time window width required for the FRFT-based linewidth is also explored. With the new decomposition thought of dynamical time-frequency lasers, the laser spectrum purity measurement resolution is no longer limited by the laser wavelength-scanning speed. In theory, the spectral linewidth resolution is inversely proportional to the measurement time window. The new understanding of the energy distribution of dynamical time-frequency laser proposed in this work will contribute to the accurate characterization of laser dynamics with fast and broadband frequency sweeping, and benefit for the optical precision measurement applications.

## 2. FRFT theory and simulation

The comparison of energy spectrum distribution between the traditional Fourier transform (FT) and fractional Fourier transform (FRFT) is shown in Figure 1. For a fixed single-frequency laser signal, the traditional FT can accurately analyze the frequency domain information of the signal, as shown in Figure 1(a). As the center frequency remains constant, the decomposition of the signal to the fixed-frequency components will not cause the broadening of the spectrum linewidth. However, for a dynamical time-frequency laser signal in a certain time, such as the wavelength-swept laser, the fixed-frequency decomposition of the laser signal will lead to the broadening of the spectrum bandwidth, as shown in Figure 1(b). Different from the traditional FT, the decomposition component of FRFT is the chirp-frequency waveform signal, which can be coincided with the wavelength-swept laser signal, and is promising to obtain a spectrum-concentrated energy distribution, as shown in Figure 1(c). In the demodulation process based on FRFT, the frequency sweep rate can be tuned by choosing different rotation angles (or fractional domain order), in order to obtain the narrowest energy spectrum distribution for wavelength-swept lasers.

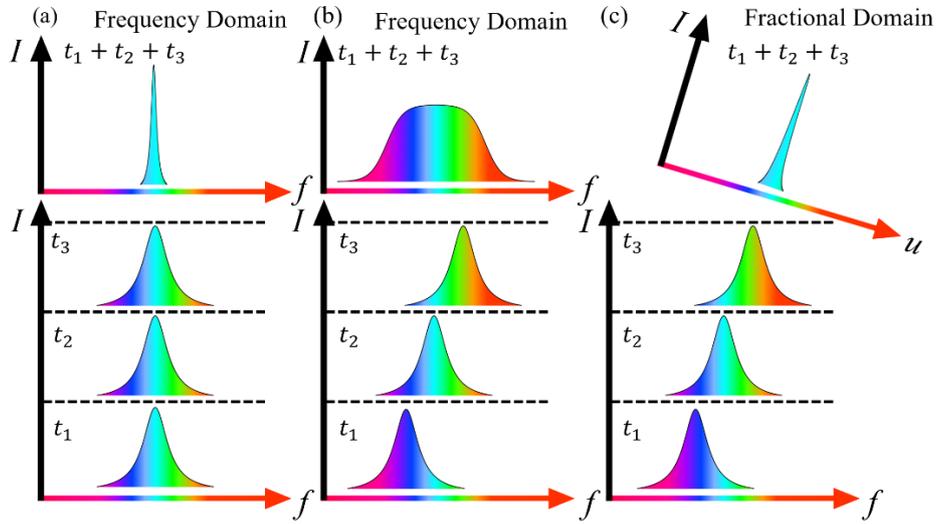

**Figure 1.** The comparison of energy spectrum distribution between the traditional Fourier transform (FT) and fractional Fourier transform (FRFT). (a) The energy distribution by traditional FT tends to be of single narrow band with increased observation time, for a fixed single-frequency signal. (b) Wide-band broadening occurs with increased observation time of traditional FT, for a frequency-swept signal. (c) FRFT can still concentrate the energy distribution span of the frequency-swept signal with increased observation time.

The FRFT[22,23] is a generalization of the Fourier transform with additional free angle parameters. It can be interpreted as a rotation by an angle $\alpha$ in the time-frequency plane[21]. The FRFT of a function $f(t) \in L^2(\mathfrak{R})$ is defined as[22]

$$F_\alpha(u) = \mathrm{F}^\alpha \{f(t)\}(u) \triangleq \int_\mathbb{R} f(t) \mathrm{K}_\alpha(u,t) \mathrm{d}t , \tag{1}$$

where $\mathrm{F}^\alpha$ denotes the FRFT operator, $\alpha$ is the rotation angle of time-frequency plane, $u$ denotes

the fractional-domain frequency, and the kernel function $K_\alpha(u,t)$ is given by

$$K_\alpha(u,t) = \begin{cases} A_\alpha e^{j\frac{u^2+t^2}{2}\cot\alpha - jtu\csc\alpha}, & \alpha \neq m\pi \\ \delta(t-u), & \alpha = 2m\pi \\ \delta(t+u), & \alpha = (2m-1)\pi \end{cases} \tag{2}$$

where $A_\alpha = \sqrt{(1-j\cot\alpha)/2\pi}$, $\delta$ is the delta function, and $m \in \mathbb{Z}$. It is worth noting that $\mathrm{F}^{2\pi m}$ is the identity operator for any integer $m$, and $\mathrm{F}^\alpha$ has the property of $\mathrm{F}^{\alpha+\beta}\{f(t)\} = \mathrm{F}^\alpha\{\mathrm{F}^\beta\{f(t)\}\}$. For $\alpha = \pi/2$, Eq. (1) is reduced to the traditional Fourier transform. The inverse FRFT is given by

$$f(t) = \mathrm{F}^{-\alpha}\{F_\alpha(u)\} \triangleq \int_{-\infty}^{+\infty} F_\alpha(u) K_\alpha^*(u,t) \mathrm{d}u, \tag{3}$$

which means that the inverse FRFT can be regarded as the FRFT with rotation angle $-\alpha$.

We can assume that a typical wavelength-swept laser signal to be analyzed satisfies the form of $e^{j\frac{k_0}{2}t^2 + \omega_0 t + \varphi_0(t)}$, where $k_0$ is the frequency tuning rate, $\omega_0$ is the initial frequency, and $\varphi_0(t)$ is the noise phase. When $\alpha$ is specially chosen so that $\cot\alpha = -k_0$, the quadratic factor $e^{+j\frac{\cot\alpha}{2}t^2}$ of FRFT can accurately offset the frequency sweep of the signal, and the linear factor $e^{-jtu\csc\alpha}$ of FRFT is equivalent to a traditional FT of the remaining phase part $e^{j(\omega_0 t + \varphi_0(t))}$, with the variable $\omega = u\csc\alpha$ as the Fourier frequency. Therefore, the laser linewidth of the wavelength-swept laser, which can reflect the laser noise $\varphi_0(t)$, can be defined as

$$\Delta f = \frac{1}{2\pi}\Delta u \csc\alpha, \tag{4}$$

where $\Delta u$ is the half energy spectral width of the laser signal decomposed by FRFT at $\alpha$ rotation angle, which satisfies $\int_{u_0-\Delta u/2}^{u_0+\Delta u/2} F_\alpha(u) F_\alpha^*(u) \mathrm{d}u = \frac{1}{2}\int_{-\infty}^{+\infty} F_\alpha(u) F_\alpha^*(u) \mathrm{d}u$, where $u_0$ is the fractional domain center frequency, defined as $u_0 = \int_{-\infty}^{+\infty} u F_\alpha(u) F_\alpha^*(u) \mathrm{d}u / \int_{-\infty}^{+\infty} F_\alpha(u) F_\alpha^*(u) \mathrm{d}u$.

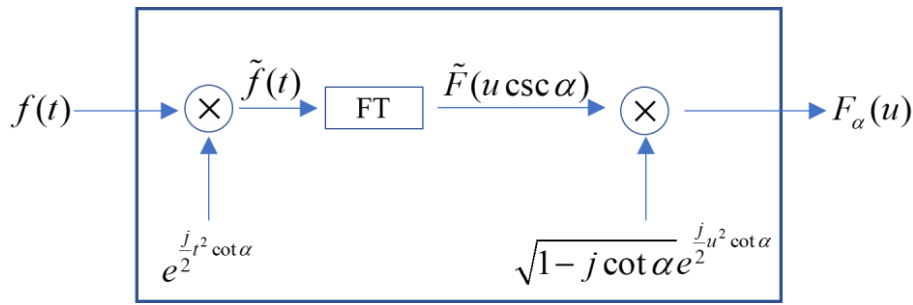

**Figure 2.** Decomposition structure diagram of the FRFT.

For calculation of FRFT, T. Erseghe et al.[26] propose an FRFT decomposition structure as show in Figure 2. The core of the algorithm is FT, and two chirp operators are used for the input signal and output signal. We simulate the energy distribution of wavelength-swept laser signal based on FRFT, as shown in Figure 3. Based on the basic properties of FRFT, the rotation angle $\alpha$ represents the angular frequency sweep speed $k$ of the chirp component in the inverse FRFT, with $k = -\cot\alpha$. When the rotation angle $\alpha$ is scanned from 0 to $2\pi$, the energy distribution can be concentrated at a specific angle, as shown in Figure 3(a). Specially, the spectrum distribution with a typical rotation angle is shown in Figure 3(b). From the half energy bandwidth $\Delta u$ of the FRFT, we can get the linewidth of swept laser, which can be varied by the rotation angle. Therefore, for the narrowest measurement of linewidth, one should search for the optimal rotation angle.

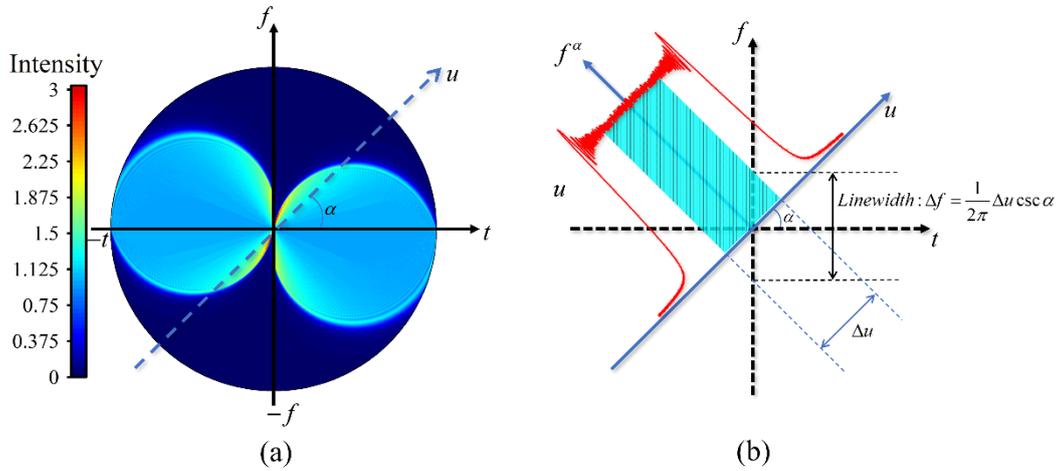

**Figure 3.** Simulation of fractional Fourier transform with different rotation angles. (a) The intensity distribution of FRFT with different rotation angles. (b) The definition of linewidth with a specific rotation angle.

## 3. Coherent detection configuration

The delayed self-heterodyne coherent detection system is utilized to measure the transient frequency and phase of the wavelength-swept laser, as shown in Figure 4. The system consists of two couplers $C_1$ and $C_2$, an acousto-optic modulator (AOM), an AOM driver, a 10 m delay fiber, a photodetector (PD), a data acquisition system and the computer. The AOM provides a frequency shift for the laser signal, and the delay fiber provides a time delay $\tau$ for the detection system.

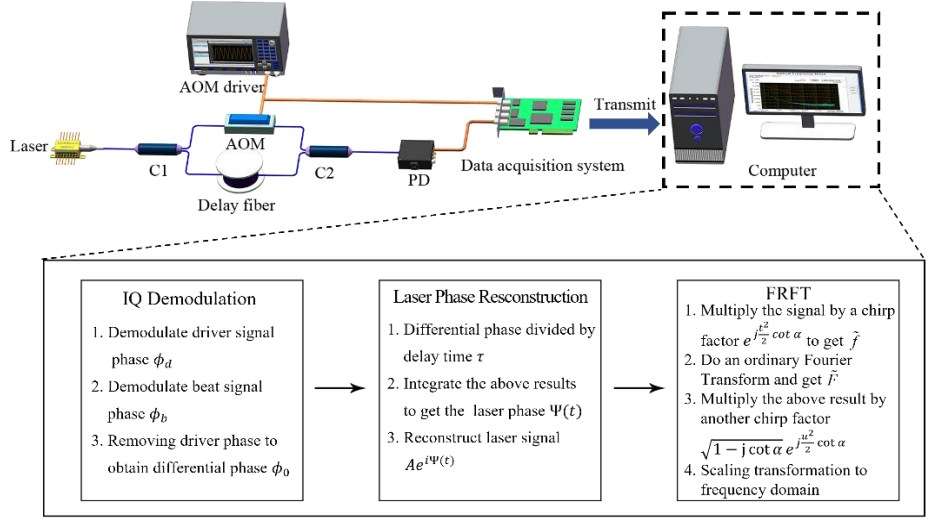

**Figure 4.** The delayed self-heterodyne coherent detection system and data analysis process. $C_1$ and $C_2$: couplers; AOM: acousto-optic modulator; PD: photo detector.

Based on the delayed self-heterodyne coherent detection system, the wavelength-swept laser signal can be expressed as

$$E = Ae^{i\left(\omega t+\frac{k}{2}t^2+\varphi(t)\right)}, \tag{5}$$

where $A$ is the laser amplitude, $\omega$ is the initial frequency, $k$ is the sweep rate, and $\varphi(t)$ is the noise phase. The laser signal passing through the AOM can be expressed as

$$E_1 = A_1 e^{i\left((\omega+\omega_0)t+\frac{k}{2}t^2+\varphi(t)\right)}, \tag{6}$$

where $A_1$ is the laser amplitude through the AOM, and $\omega_0$ is the shifted frequency by AOM. The laser signal passing through delay fiber can be expressed as

$$E_2 = A_2 e^{i\left(\omega(t-\tau)+\frac{k}{2}(t-\tau)^2+\varphi((t-\tau))\right)}, \tag{7}$$

where $A_2$ is the laser amplitude through the delay fiber, and $\tau$ is the delay time. The beat signal obtained at coupler $C_2$ can be expressed as

$$I = |E_1+E_2|^2 = A_1^2 + A_2^2 + 2A_1A_2\cos\left(\omega_0 t+\omega\tau+kt\tau-\frac{k}{2}\tau^2+\varphi(t)-\varphi(t-\tau)\right). \tag{8}$$

By demodulating the signal, we can get the phase $\phi_b(t)$ of beat signal, i.e. $\phi_b(t)=\omega_0 t+\omega\tau+kt\tau-\frac{k}{2}\tau^2+\varphi(t)-\varphi(t-\tau)$. In the meantime, the phase $\phi_d(t)=\omega_0 t$ of the RF driving signal of AOM can also be demodulated according to the synchronously recorded waveform. Therefore, the demodulated laser phase noise related item $\phi_0(t)$ satisfies

$$\phi_0(t) = \phi_b(t) - \phi_d(t) = k\tau t + \varphi(t) - \varphi(t-\tau) + \omega\tau - \frac{k}{2}\tau^2, \quad (9)$$

where $k\tau$ could be regarded as the linear fitting slope of the demodulated phase $\phi_0(t)$ over time, $\varphi(t) - \varphi(t-\tau)$ denotes the residual noise after the linear fitting of $\phi_0(t)$ over time, and $\omega\tau - k\tau^2/2$ denotes the initial phase of $\phi_0(t)$. Under the condition of short time delay $\tau$, the derivative of phase noise could be written as

$$\frac{d\varphi}{dt} \approx \frac{\varphi(t) - \varphi(t-\tau)}{\tau} = \frac{\phi_0(t)}{\tau} - kt - \omega + \frac{k\tau}{2}. \quad (10)$$

The wavelength-swept laser phase $\Psi(t)$ can be reconstructed as

$$\Psi(t) = \omega t + \frac{k}{2}t^2 + \varphi(t) = \int (\omega + kt + \frac{d\varphi}{dt})dt$$
$$\approx \int (\frac{\phi_0(t)}{\tau} + \frac{k\tau}{2})dt = \frac{1}{\tau}\int (\phi_0(t) + \frac{k\tau^2}{2})dt$$
$$= \frac{1}{\tau}\int \phi_0(t)dt, \quad (11)$$

where the initial phase of $\phi_0(t)$ is hard to demodulate, which has to be set as an arbitrary value, and will induce an arbitrary initial frequency offset of the wavelength-swept laser. Based on the algorithm shown in Figure 2, we can further obtain the FRFT of the reconstructed wavelength-swept laser as

$$F_\alpha(u) = \sqrt{1 - j\cot\alpha}\, e^{j\frac{u^2}{2}\cot\alpha}$$
$$\frac{1}{\sqrt{2\pi}}\int_{-\infty}^{+\infty} \left( A e^{i\Psi(t)} e^{j\frac{t^2}{2}\cot\alpha} \right) e^{-jtu\csc\alpha} dt, \quad (12)$$

where $u$ is the fractional frequency, $\alpha$ is the rotation angle. As mentioned above, when the rotation angle is tuned, the fractional domain spectrum will show different energy concentration, based on which we can judge the best fractional domain order. From Eq. (9), we can obtain the estimated sweep slope $k$ of the wavelength-swept laser, and the optimal rotation angle should satisfy $k = -\cot\alpha$. After the above steps, we can demodulate the wavelength-swept laser signal to a series of chirp-frequency components, and then get the intrinsic spectrum information, which is determined by the laser phase noise.

## 4. Experimental results and discussion

As proof of concept, we measure the dynamical parameters of a commercial wavelength-swept laser (Luna Phoenix 1202) as the typical dynamical time-frequency laser. In the experiment, we reconstruct the laser phase and measure the linewidth with different sweep speeds by FRFT. For each group of signals, we first use IQ demodulation to get the differential phase $\phi_0(t) = \phi_b(t) - \phi_d(t)$ and use linear fitting to obtain the reference sweep speed $k$ related to the rotation angle $\alpha$, as shown in Figure 5(a). The reconstructed phase can be further obtained by integrating the differential phase, as shown in Figure 5(b). With the optimal rotation angle $\alpha$, the distribution of energy concentration is obtained by FRFT of the reconstructed laser signal, as shown in Figure 5(c). The frequency axis is rewritten in the frequency domain, with $f = u \csc\alpha / 2\pi$, which could reflect the laser spectrum induced by laser phase noise. To characterize the linewidth of the laser signal in the fractional domain, we use the half energy width as its linewidth. The area of the shaded part is half of the overall energy, and its width is the half energy width. Figure 5(d) is the integrated energy $E(\delta u)$ curve under the condition of different integration bandwidths, which is calculated as $E(\delta u) = \int_{u_0 - \delta u/2}^{u_0 + \delta u/2} F_\alpha(u) F_\alpha^*(u) \mathrm{d}u$, where $u_0$ is the fractional domain central frequency, and $\delta u$ is the fractional domain integration bandwidth. The integration bandwidth is rewritten in the form of frequency, i.e. $\delta f = \delta u \csc\alpha / 2\pi$. Considering the random error during the calculation, each signal is calculated by 10 times and averaged to obtain the linewidth.

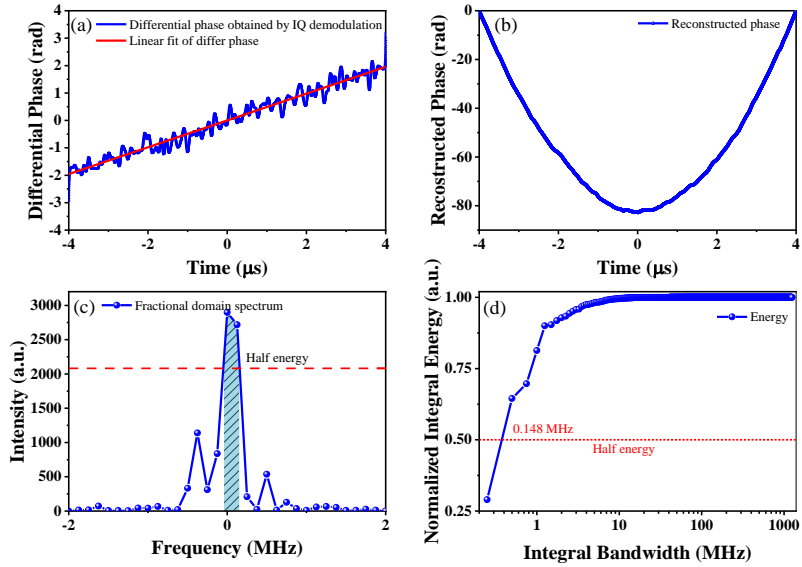

**Figure 5.** Processing flow of the FRFT algorithm (Sweep speed is 10.02 nm/s). (a) Differential phase. (b) Reconstructed laser phase. (c) Fractional domain spectrum. (d) Normalized integrated energy.

Considering that FRFT is a generalized extension of traditional Fourier transform, we explore the influence of the time window width of FRFT, as shown in Figure 6. It is worth noting that the narrowest measured linewidth could be obtained with suitable time window width. When the width of the time window is too short, due to the insufficient sampling time, the frequency analysis resolution is not enough to show the laser noise. When the width is too long, the measured linewidth is broadened due to the 1/$f$ frequency noise of the laser cavity, which has also been widely discussed in fixed single-frequency lasers[27].

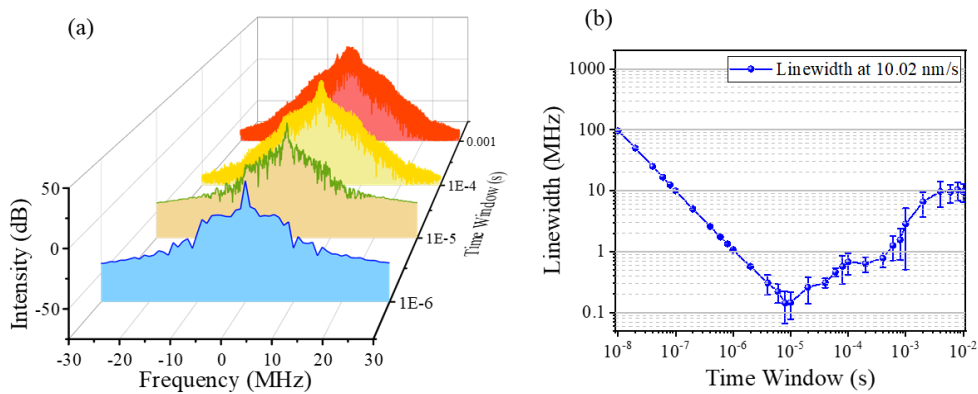

**Figure 6.** (a) Fractional domain spectrum. (b) Half energy linewidth under different time windows.

The linewidths with different sweep speeds from 0.93 nm/s to 1000.03 nm/s are further measured, as shown in Figure 7. The sweep speeds are directly set by the control program of the commercial wavelength-swept laser. Based on the reconstructed phase in Figure 7(a), the linewidths under different time windows could be measured by suitable rotation angles, as shown in Figure 7(b). The optimal time windows for different sweeping speeds are very close, and the narrowest linewidths are almost equal, which means that the sweep speed nearly does not change the intrinsic linewidth of the laser. However, when the time window is enlarged, the faster the frequency sweep speed is set, the broader the linewidth is measured, which are mainly induced by the nonlinear sweeping of the laser frequency. Therefore, the narrowest linewidth can only be measured in the appropriate time window width for wavelength-swept lasers.

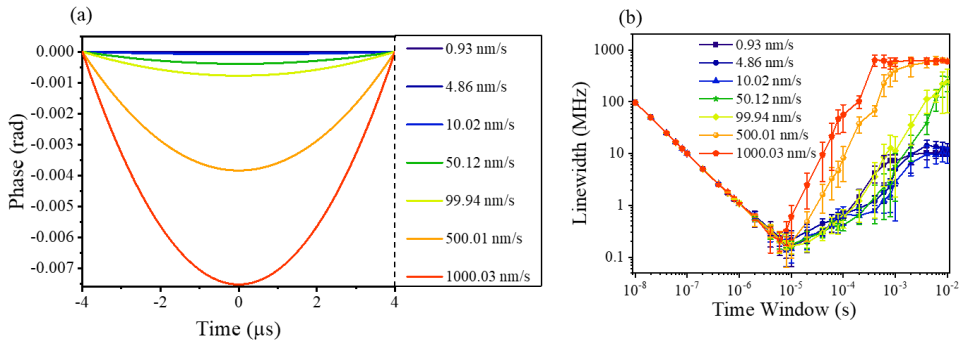

**Figure 7.** (a) Reconstructed phase with different sweep speeds. (b) The measured linewidths with different sweep speeds in different time windows.

To show the influence of rotational angle for the FRFT of the reconstructed laser signals, we set up different fractional orders to measure the laser linewidth at a same sweep speed of 10.02 nm/s, as shown in Figure 8(a). The front five curves are the fractional domain spectra near the optimal rotational angle, and the last one is the spectrum with $\alpha=\pi/2$, i.e. the traditional FT spectrum. It is worth noting that the deviation from the optimal rotation angle will induce the spectral broadening. Specially, we further compare the spectra by FRFT and FT in different time window widths, as shown in Figure 8(b). The narrowest linewidths calculated by FRFT and FT are 0.14 MHz and 0.84 MHz, respectively. The results show that the measured linewidth by traditional FT, with the fixed single-frequency waveforms as the decomposition components, will be largely broadened due to the high-speed frequency sweep rate of the laser. Meanwhile, the measured narrowest linewidth by FRFT remains relatively constant, with the chirp-frequency waveform as the decomposition components, which demonstrates that the FRFT can break through the measurement resolution limitation of traditional FT and has higher accuracy in the decomposition of wavelength-swept laser signal.

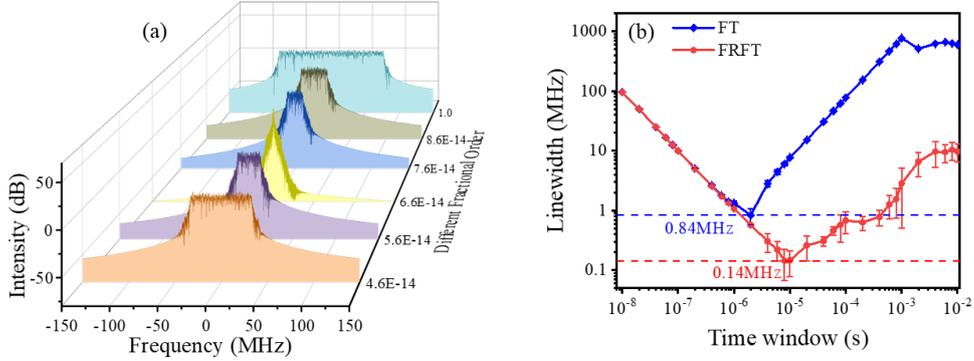

**Figure 8.** (a) Fractional domain spectra of experimental data under different fractional orders. (b) Comparison of linewidths between the fractional Fourier transform and traditional Fourier transform of experimental data.

## 5. Conclusion

In summary, we propose a generalized concept of wavelength-swept laser linewidth considering the essence of chirp-frequency signal, which is realized by fractional Fourier transform (FRFT) of the coherently reconstructed laser waveform. In the proof-of-concept experiment, the narrow linewidth of a commercial wavelength swept laser is tested with different scanning speeds from 1 nm/s to 1000 nm/s. The sweep speed and instantaneous spectrum can be reconstructed simultaneously. The optimal demodulation time window of signals with different sweep speeds is also verified. In theory, the spectral linewidth resolution is inversely proportional to the measurement time window. However, in the actual measurement, the noise also increases due to the increase of resolution. Therefore, the choice of time window is also very important. With the new decomposition thought of wavelength-swept lasers, the new understanding of the energy distribution of wavelength-swept laser proposed in this work will contribute to the dynamic and accurate characterization of wavelength-swept laser parameters, and benefit for the optical precision measurement applications.

**Acknowledgements**

National Natural Science Foundation of China (NSFC) (61927818, 61635004, 61935007,